\documentclass[iop]{emulateapj}

\usepackage{amsmath}
\usepackage{natbib}
\usepackage{color}
\usepackage{hyperref}
\hypersetup{colorlinks=true,urlcolor=black,linkcolor=red,citecolor=blue}
\newcommand{\msun}{M_\odot}

\shorttitle{Signatures of AGN Jet Triggered Star Formation}
\shortauthors{Dugan et al.}

\begin{document}

\title{Stellar Signatures of AGN Jet Triggered Star Formation}
\author{Zachary Dugan\altaffilmark{1}, Sarah Bryan\altaffilmark{2}, Volker Gaibler\altaffilmark{3}, Joseph Silk\altaffilmark{1,4,5}, Marcel Haas\altaffilmark{6}}
\altaffiltext{1}{The Johns Hopkins University Department of Physics \& Astronomy, Bloomberg Center for Physics and Astronomy, Room 366
3400 N. Charles Street, Baltimore, MD 21218, USA}
\altaffiltext{2}{The University of Manchester, School of Physics \& Astronomy, Jodrell Bank Centre for Astrophysics, Alan Turing Building, Oxford Road, Manchester, M13 9PL, United Kingdom}
\altaffiltext{3}{Universit\"at Heidelberg, Zentrum f\"ur Astronomie, Institut f\"ur Theoretische Astrophysik, Albert-Ueberle-Str. 2, 69120 Heidelberg, Germany}
\altaffiltext{4}{Beecroft Institute for Cosmology and Particle Astrophysics, University of Oxford, Keble Road, Oxford OX1 3RH, UK}
\altaffiltext{5}{Institut d'Astrophysique de Paris, UMR 7095, CNRS, UPMC Univ. Paris VI, 98 bis Boulevard Arago, 75014 Paris, France}
\altaffiltext{6}{Rutgers University Department of Physics \& Astronomy, 136 Frelinghuysen Rd, Piscataway, NJ 08854}

%\keywords{ galaxies: star formation--- galaxies: active---
%galaxy: formation---galaxies: evolution}

%%%%%%%%%%%%%%%%%%%%%%%%%%%%%%%%%%%%%%%%%%%%%%%%%%%%%%%%%%
\begin{abstract} 
To investigate feedback between relativistic jets emanating from Active Galactic Nuclei (AGN) and the stellar population of the host galaxy, 
we analyze the long-term evolution of the galaxy-scale simulations by \citet{Gaibler12} of jets in massive, gas-rich galaxies at $z\sim 2\mbox{--}3$ and of 
stars formed in the host galaxies.  We find strong, jet-induced differences in the resulting stellar populations of galaxies that host relativistic jets and 
galaxies that do not, including correlations in stellar locations, velocities, and ages.  
Jets are found to generate distributions of increased radial and vertical velocities that persist long enough to effectively extend the stellar 
structure of the host. 
The jets cause the formation of bow shocks that move out through the disk, generating rings of star formation 
within the disk. The bow shock often accelerates pockets of gas in which stars form, yielding populations of stars with significant radial 
and vertical velocities, some of which have large enough velocities to escape the galaxy.  These stellar population signatures can serve to 
identify past jet activity as well as jet-induced star formation.    
\end{abstract}

%%%%%%%%%%%%%%%%%%%%%%%%%%%%%%%%%%%%%%%%%%%%%%%%%%%%%%%%%%
\section{Introduction}

The impact that Active Galactic Nuclei (AGN) have on their hosts is one of the most critical components of galaxy formation theory \citep{Silk98}.  Observationally, astronomers find fewer luminous galaxies than expected based on the evolution of a $\Lambda$CDM universe and frequently attribute this phenomenon to AGN quenching of star formation \citep{Weinmann}, ultimately reducing the number of bright galaxies. Some cosmological simulations use the injection of thermal energy or mechanical heating into the center of galaxies as a prescription for AGN feedback \citep[e.g.][]{Springel,Sijacki}, and often derive luminosity functions that match up better with observations. However, the physical processes in AGN feedback are far more complex than the introduction of heating processes.  Recent computational analyses of jet-driven feedback shows that relativistic jets can actually induce star formation in their host galaxies \citep{Antonuccio-Delogu,Gaibler12}. 

AGN jets have largely been assumed to reduce star formation rates because the jets increase the temperature of the gas they collide with, making collapse under self-gravity more difficult.  Interestingly, both recent simulations and observations reveal that jets can actually foster star formation by creating some regions of high density and low temperature embedded in the cocoon surrounding the jet \citep{Antonuccio-Delogu}. Expanding jets create cocoons of turbulent gas surrounding the jet, and the overpressured cocoons form bow shocks and hit clouds of cold gas, increasing density and creating regions of star formation \citep{SilkMamon12}. The bow shock generated from jets can also potentially collapse pockets of cold gas to form stars as the cocoon expands out along the disk and compresses it.  For example, both Minkowski's object and Hanny's Voorwerp object are galaxy-scale gas-dominated clouds struck by  jets that show high star formation rates \citep{Croft,Rampadarath}.  Other observations from \citet{Juneau13} and \citet{Fang12} also show extended star formation in galaxies that host AGN, while \citet{Rauch} even finds star formation potentially triggered by an AGN at a redshift of $z=3.045$. \citet{Ishibashi12} also provides a theoretical framework for AGN triggered star formation.  
        
Unfortunately, AGN jet-triggered star formation is difficult to observe because AGN jets are short-lived on a cosmological time scale ($\sim 10^7$ yr), which makes simulations critical to comprehending this phenomenon.  Furthermore, the difficulty of simultaneously observing jets and star formation in the host may obscure the frequency of positive jet feedback, as discussed in \citet{Zubovas13}. Once a more complex, more accurate relationship between jets and star formation is established, previously unexpected characteristics of stellar populations of galaxies at high redshifts may be explicable.  In this paper, we aim to find the impacts of jets on host galaxy stellar populations that endure substantially longer than the jet itself to further help establish the role of jets in galaxy formation and evolution.  We analyze the hydrodynamic simulations by \citet{Gaibler12} of jets in a gas rich, $z \sim 2$ galaxy, and then integrate the orbits of stars formed during the simulation for a Gyr to find signatures of past AGN-triggered star formation. 
AGN feedback may also link to the evolution of the host galaxy's morphology.  Several studies show spherical and cylindrical symmetry in morphological effects generated from  AGN interactions \citep{Fang12, Liu}.   Observations and arguments from \citet{Ishibashi}, \citet{Bezanson}, \citet{van Dokkum}, and \citet{Patel} show that galaxies of a particular mass at $z \sim 2$ are more dense than galaxies of roughly the same mass at a redshift of $z \sim 0$, indicating that a mechanism, presumable mergers or AGN feedback, is expanding the size of these galaxies over a period of roughly 10 Gyr.  Similar observations show that the central regions of the galaxies at $z \sim 2$ are similar to galaxies of the same mass today, indicating that the growth in size is occurring in the outer regions, a phenomenon we see in our simulation.    We examine radial velocity distributions in our simulations to examine their impact on the growth in effective size of massive galaxies at $z \sim 2$.  
    
In this respect, AGN jets may be linked to the ongoing mystery of the origin and cause of hypervelocity stars (HVS) in our own galaxy.  Analytical arguments have been made supporting the idea that an AGN jet could be the mechanism generating high velocity stars that may or may not escape their host galaxy \citep{Silk12}. The theory is that the bow shock of the jet can both condense and accelerate pockets of gas that will then form stars which share the host cloud's velocity.  Obviously observing such a phenomenon would be difficult.  However, both momentum and energy based arguments can be made supporting the idea, as in \citet{Silk12}.  To add to this theoretical argument, we analyze stars formed in the simulation that eventually escape the host galaxy.  If jets do, in fact, induce star formation, and some of these stars are hypervelocity stars, then these high velocity stars may be excellent tracers of past jet activity and corresponding induced star formation.  Many HVS orbits appear to emanate 
from the center of our galaxy, largely contradicting the possibility of supernovae and birth kicks  as the main generator of HVS \citep{Brown12}.  We analyze our simulations to examine AGN as a possible mechanism for the generation of HVS and escape stars, which in turn could be signatures of AGN-triggered star formation.   
    
While the Milky Way is very different than the high redshift gas-rich galaxies simulated in \citet{Gaibler12} and analyzed in this paper, patterns and distributions of escaped stars from \citet{Gaibler12} could potentially also be applied to our own galaxy despite the important differences in the respective physical parameters.  Several studies indicate possible jet activity in the Milky Way's past, as in \citet{Efremov} and \citet{Yusef-Zadeh}.  The presence of the Fermi bubbles further suggests past AGN activity roughly 10 Myr ago \citep{Su,Guo,Zubovas}. Furthermore, present data on hypervelocity B stars may indicate a possible AGN outburst $\sim$100 Myr ago \citep{Brown12}.  If these time scales are indeed accurate, then the time scales of  orbit evolution should suffice to extract signs of past jet activity, jet-driven escaped stars, and perhaps jet-induced star formation. Particularly in light of the upcoming Gaia space astrometry mission, which should produce the positions and velocities of a billion stars in the Milky Way, further analysis of the jet's impact on stellar positional and velocity distributions as well as galaxy morphology should provide signatures not just of past jet activity, but also of jet-induced star formation triggering in our own galaxy.

%%%%%%%%%%%%%%%%%%%%%%%%%%%%%%%%%%%%%%%%%%%%%%%%%%%%%%%%%%
\section{Methodology}
\label{sec:method}

We analyze the four simulations of \citet{Gaibler12} examining AGN jet activity in a massive, gas-rich disk galaxy with an exponential radial profile with a 5 kpc scale radius, a scale height of 1.5 kpc, and a mass of $1.5\times10^{11}\msun$.  Gas in the disk initially has a radial and vertical profile
\begin{equation} 
\rho( \vec{x} ) \propto \exp \left\{ -r / r_0 \right\} \, \mathrm{sech}^2(h/h_0)
\end{equation}
on top of a fractal cube that mimics the clumpy interstellar medium.
The simulation utilizes the RAMSES adaptive mesh refinement code from \citet{Teyssier} with maximum resolution of grid cells that are 62.5 pc on a side.  Gravity was not included due to the short time scales in the simulation compared to the disk evolution time scale. Control simulations without a jet were performed to examine the evolution of the disk to isolate the effects of the jet on star formation and gas dynamics. In total, four simulations are performed: disk only as well as disk and jet for both low and high thresholds for star formation.
    
The simulation uses a star formation model prescribed by \citet{Rasera} which reproduces the Kennicutt--Schmidt relation \citep{Kennicutt1998}. Stars are created only in regions where the number density of hydrogen is $n_H > n_\star$, with a rate controlled by a fixed star formation efficiency value and the local free-fall time
\begin{equation}    
\dot{\rho}_\star =  \epsilon \rho / t_\mathrm{ff}  \,
\end{equation}    
where $\epsilon$ is the star formation efficiency, $t_\mathrm{ff}$ is the local free-fall time, and $n_\star$ is the star formation threshold.  This simulation looks at two cases for star formation, the first with $\epsilon=0.025$ and $n_\star=0.1$ cm$^{-3}$, and the second with $\epsilon=0.05$ and $n_\star=5$ cm$^{-3}$.  In this paper, we examine the positions, velocities, and time of formation of all the stars formed in each of the four simulations.  

For this study, we assume a gravitational potential that is generated by an exponential thin disk and a  NFW \citep{NFW1996} dark matter halo. We assume that this potential remains constant during our simulated time range. Since the mass of jet-induced stars is considerably smaller than the mass of the old, pre-existing stellar component and the dark matter halo, the change in the potential will be small. Because the hydrodynamic simulation did not include the rotation of the galaxy, a rotational velocity component must be added to each star to properly model the long-term evolution of the stellar population. This circular velocity is determined from the gravitational potential, and the rotational velocity resulting from an exponential thin disk is \citep{Mo}
\begin{eqnarray}
 v_\mathrm{c,disk}^2(r) =  -4\pi G\Sigma_0 r_0  \left( \frac{r}{2r_0} \right)^2  \times \nonumber \\
\left(I_0\left( \frac{r}{2r_0} \right)K_0\left( \frac{r}{2r_0} \right)  - I_1\left( \frac{r}{2r_0} \right) K_1\left( \frac{r}{2r_0} \right) \right) 
\end{eqnarray}
where $I_n$ and $K_n$ are the modified Bessel functions of the first and second kind, and $\Sigma_0 = M_\mathrm{b} / 2 \pi R_\mathrm{d}^2 = 0.199$ g cm$^{-2}$ is the central surface density, calculated with $M_\mathrm{b} = $ the baryonic mass in the simulation.

The simulations by \citet{Gaibler12} prescribe the total baryonic mass in the galaxy, and from this baryonic mass we determine the mass of the dark matter halo using a calibration from \citet{Anderson},
\begin{equation}    
f_\mathrm{b} = 0.04 \left( \frac{M}{2\times 10^{12} \msun} \right)^{1/2} = M_\mathrm{b} / M_\mathrm{T} \, ,
\end{equation}    
where $M_b$ is the baryon mass, and $M_\mathrm{T}$ is the total mass, equal to the sum of the baryon mass and the dark matter mass.  The simulation begins with $M_\mathrm{b} = 1.5\times10^{11}\msun$, resulting in a dark matter mass,  
$M_\mathrm{DM} = 2.89\times10^{12}\msun$. 
The NFW potential from \citet{NFW1996} is:  
\begin{equation}    
\Phi(r) =  \frac{-G M_{200} \, \mathrm{ln}(1 + \frac{r}{r_\mathrm{s}})}{r \: f(c)} 
\end{equation}    
where $M_{200}$ is the virialized mass, $r_s$ is the scale radius of the dark matter halo, c is the concentration parameter, and
\begin{equation}    
f(c)=\mathrm{ln}(1+c) - \frac{c}{1+c}  \, .
\end{equation}
We set $M_{200}$ equal to the dark matter mass $M_{DM}$, calculated above, and $r_s=50$ kpc.  We use a mass-concentration relationship from \citet{Munoz} to determine a reasonable value for $c$ given the mass of the DM halo, and determine $c=5.623$.
The circular velocity resulting from the NFW potential is therefore
\begin{equation}    
v_\mathrm{c,NFW}^2 =  \frac{-GM_{200}}{f(c)}\times \frac{r-(r+r_s) \ln \left( \frac{r+r_s}{r_s} \right)}{r(r+r_s)}  \, .
\end{equation}    

To combine the effects of the different potentials, we take advantage of the spherical symmetry and add the two rotational velocities resulting from the disk and from the dark matter together in quadrature as in \citet{Mo}.  
Because we have no knowledge of where, in the direction perpendicular to the plane of the disk, each star is in its theoretical orbit, we assume that upon the addition of the circular velocity, each star is at its maximum height above or below the disk.  As such, at that point in the star's orbit, the component of the star's velocity perpendicular to the plane of the disk is zero.  Therefor, as we break down the centripetal velocity to be added to the star's initial velocity, we break it down only into the components in the plane of the galaxy's disk, the y-z plane.  

We then integrate all the orbits for 1 Gyr after the simulation ends with a Runge-Kutta-Fehlberg routine.  We examine ten snapshots between 0 and 0.1 Gyr and ten snapshots between 0.1 and 1 Gyr in order to analyze the evolution of the galactic morphology as well as various velocity and parameter distributions.  We calculate the following parameters for the snapshots from 0 to 1 Gyr: the spin parameter $\lambda$, the velocity anisotropy parameter $\beta$, and the shape parameters $s$, $T$, and $b$.  The spin parameter is calculated using the \citet{Bullock01} definition
\begin{equation}    
\lambda^{'} = \frac{J}{\sqrt{2} \, M v R} \, ,
\end{equation}    
where $M$ is the mass contained within a radius $R$, and $J$ is the angular momentum within the same radius.  $v$ is the circular velocity as a function of radius $R$, calculated as $v^2=GM/R$.  
As a quantification of the proportion of radial to tangential orbits, we also calculate the velocity anisotropy parameter \citep{Binney}:
\begin{equation}    
\beta = 1 - 0.5\frac{ \sigma_\mathrm{t}^2}{ \sigma_\mathrm{r}^2}
\end{equation}    
where the radial velocity dispersion is represented by $\sigma_\mathrm{r}$ the tangential velocity dispersion by $\sigma_\mathrm{t}$.  A galaxy with a $\beta=1$ is composed of purely radial orbits, and galaxies with more tangential motion have more negative values.  

In order to quantify the shape and evolving morphology of the galaxy, we begin with the calculation of the inertial tensor as in \citet{Bryan13}: 
\begin{equation}    
I_\mathrm{ij} =  \sum_k \frac{ r_\mathrm{k,i}r_\mathrm{k,j} }{r_\mathrm{k}^2}
\end{equation}
We then diagonalize the inertia tensor and calculate the eigenvalues and eigenvectors. We define the parameters $a$, $b$, $c$ as the square roots of the eigenvalues (where $a \ge b \ge c$).  Finally we compute the shape parameter $s$, which quantifies how spherical a galaxy is, as $s = c/a$.  A distribution that is perfectly spherical will have $s = 1$.  To quantify triaxiality, we compute the shape parameter $T = (a^2 - b^2)/(a^2 - c^2)$.  A galaxy with $T = 1$ (0) corresponds to an prolate (oblate) halo.  These shape parameters allow the quantification of the morphology of the galaxy, which in turn facilitates analyzing the evolution of the galactic distribution of stars as a direct result of the AGN jet. It should be noted that only stars formed in the hydrodynamic simulation (10 Myr without jet, plus $\sim 15$ with active jet) are taken into account. Since the pre-existing stellar population is not affected by the jet activity, the stellar population properties rather indicate the changes in these properties than actual measurable values. Also the feedback duration will impact the strength of these changes. Furthermore it should be kept in mind that the velocities used in this analysis taken from the hydrodynamic simulation, which by far cannot resolve the actual scales of star formation within molecular clouds. The stellar velocities, as inherited by their parent gas cloud, will certainly depend on the efficiency of momentum transfer onto the molecular gas. In the present paper, we cannot properly account for these efficiencies and rather have to interpret the velocities conservatively as upper limits.

%%%%%%%%%%%%%%%%%%%%%%%%%%%%%%%%%%%%%%%%%%%%%%%%%%%%%%%%%%
\section{Evolution of the stellar population}

We find a substantial difference between the age, velocity, and number distribution of stars created in the simulations with and without the relativistic jet, specifically after the jet has begun. The simulations with the jet show much larger vertical velocity distributions, larger azimuthal (rotational) velocity distributions, larger radial velocity distributions.  Furthermore, there are other locational signatures of jet activity, such as rings of star formation.  

\begin{figure}
\includegraphics[width=1.\linewidth,angle=0]{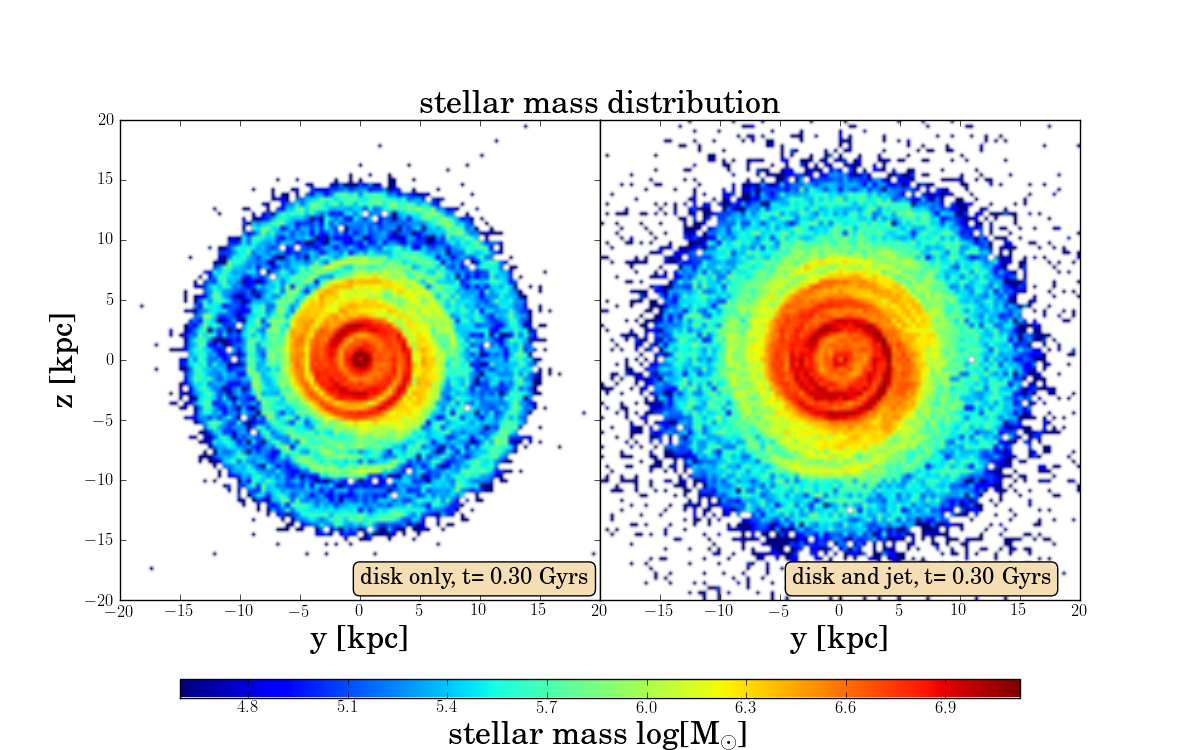}
\includegraphics[width=1.\linewidth,angle=0]{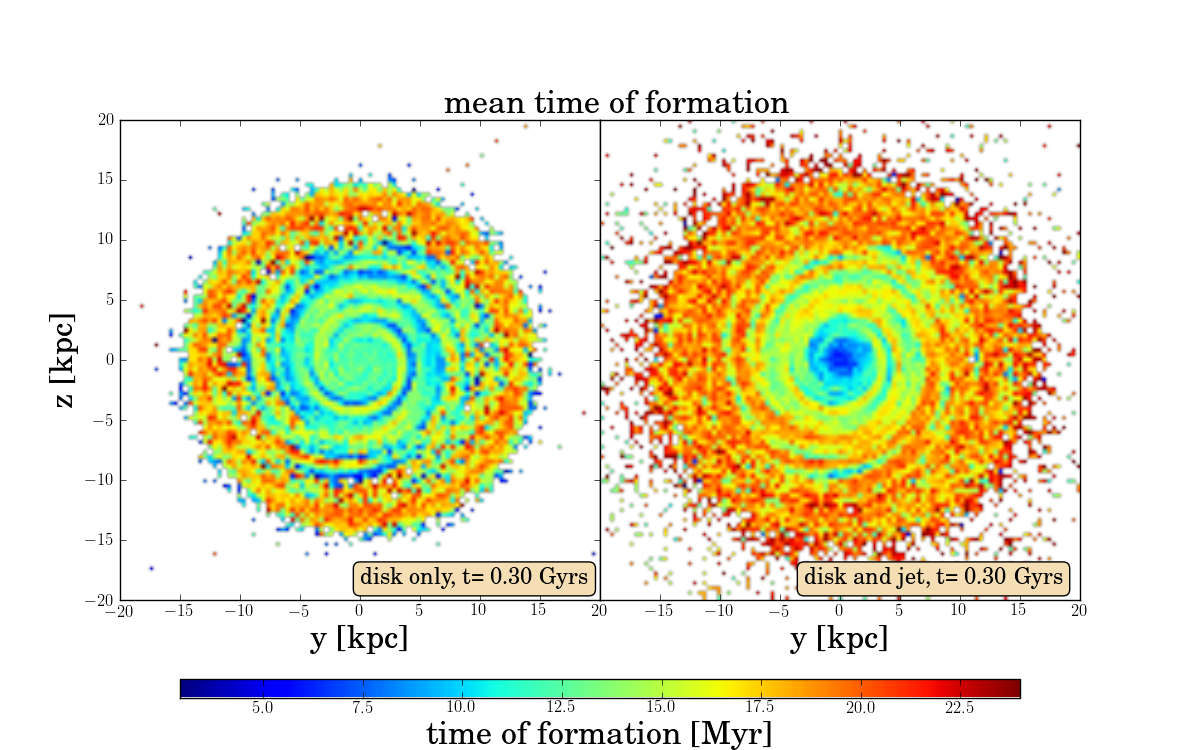}
\caption{Plots of the integrated stellar mass distribution (\emph{top}) and the mean times of formation (\emph{bottom}) at $t=0.03$ Gyr for the simulations without the jet (\emph{left}) and with the jet (\emph{right}).}
\label{fig:time_formation_stellar_mass}
\end{figure}
Figure \ref{fig:time_formation_stellar_mass} shows plots of the mean time of formation of stars within the galaxies with and without the jet.  The disk has been divided into 10,000 grid cells and color coded with the mass-weighted average time of formation of stars in each cell.  A circular cavity with a radius of $\sim 2$ kpc is carved out of the disk galaxy where the mean time of formation is below $7.5$ Myr, well before the jet is started.  In this central part of the disk, little star formation occurs after 10 Myr, when the jet starts.  The jet creates a cocoon and subsequent bowshock that re-pressurize the disk, a process that moves radially outward, creating rings of star formation along the disk. 

\begin{figure}
\includegraphics[width=1.\linewidth,angle=0]{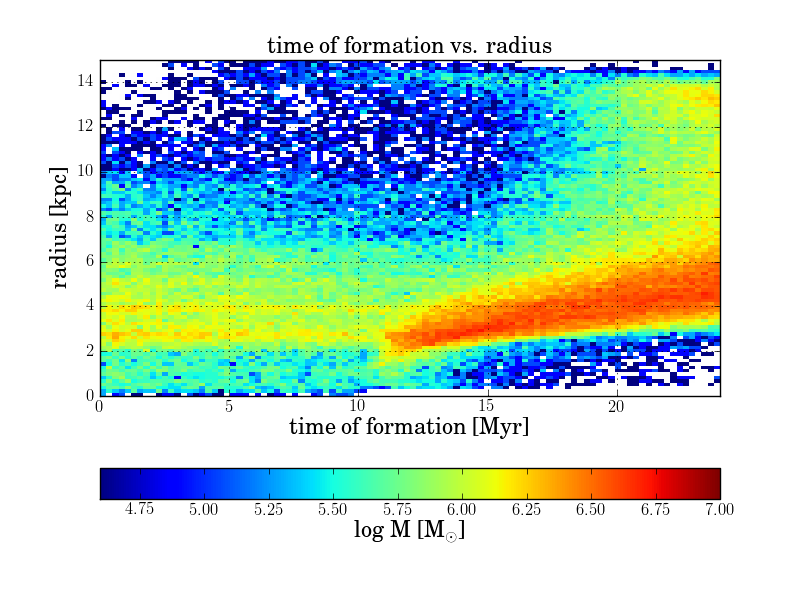}
\caption{2D histogram of stellar time of formation and the axial distance, weighted by stellar mass.}
\label{fig:phase_time_formation_radius}
\end{figure}

Figure \ref{fig:phase_time_formation_radius} is a 2D histogram of the time of formation vs. the star's distance from the center at formation, weighted by stellar mass.  The jet begins at 10 Myr, so the data before 10 Myr reflects galactic evolution without a jet.  The data after the jet begins at 10 Myr shows not only that more star formation occurs, but also where and when these stars are forming.  The feature beginning at 10 Myr and extending from 2 kpc outward radially represents a ring of star formation that begins when the jet turns on.  This ring then moves outward along the disk from a radius of 2 kpc at 10 Myr to 6 kpc at 24 Myr.  Furthermore, the data from the simulation with the jet shows additional star formation further out along the disk beyond a radius of 6 kpc all the way to a radius of 14 kpc.  Both the bow shock from the jet and the pressurization of the disk by the cocoon cause this additional star formation.  The region within a radius of $\sim$2 kpc where star formation nearly ceases shortly after 10 Myr is obvious from this plot.

\begin{figure}
\includegraphics[width=1.\linewidth,angle=0]{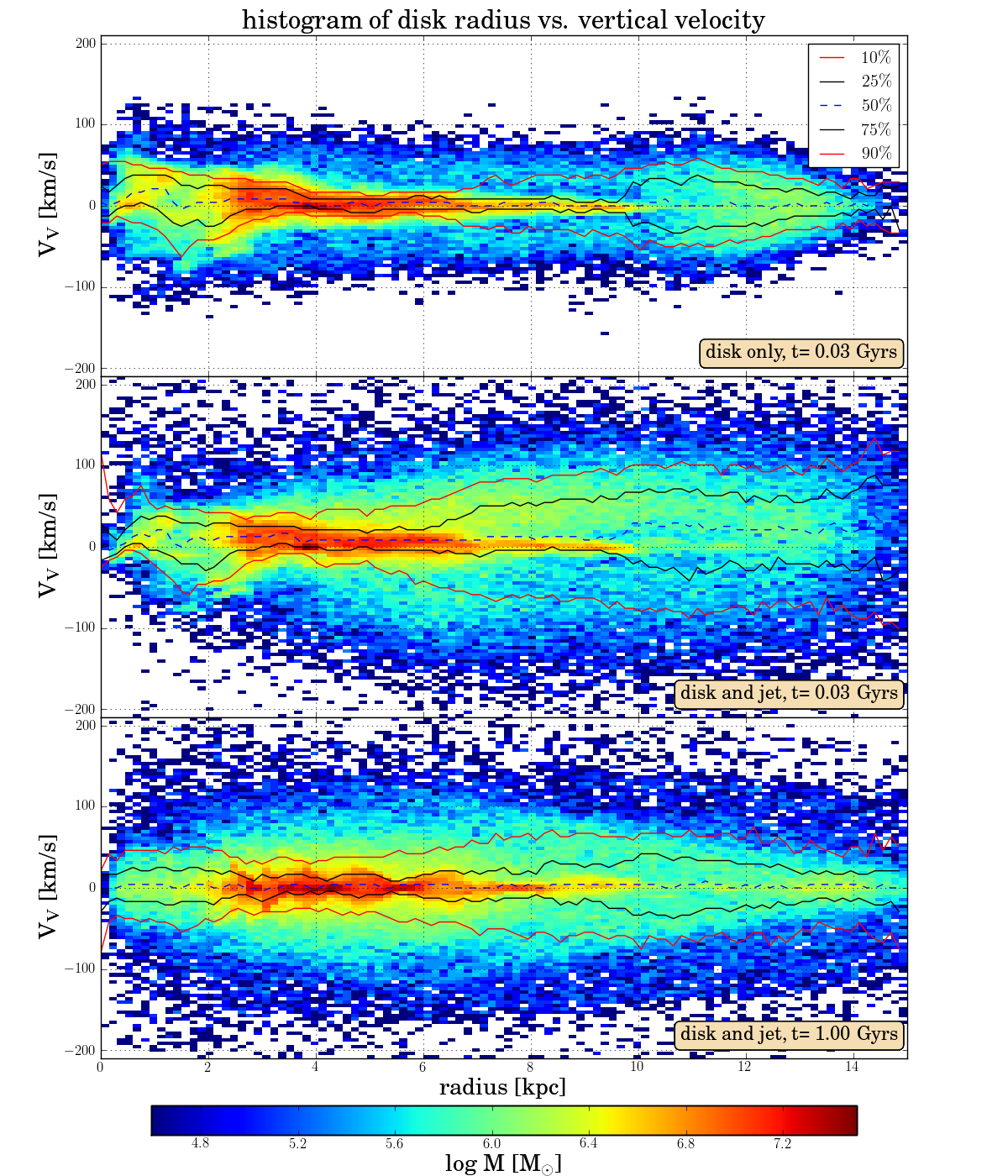}
\caption{2D histogram of vertical velocity and axial distance (cylindrical radius coordinate) for the disk only simulation at 30 Myr, the disk and jet simulation at 30 Myr, and the disk and jet simulation at 1 Gyr, weighted by stellar mass.}
\label{fig:phase_radius_vertical_velocity}
\end{figure}

We find substantial differences in stellar velocity distributions between galaxies with and without jets that endure for 1 Gyr after the beginning of the simulation, long after the jet has stopped.  The jets induce larger vertical velocities off the disk as well as larger radial velocities along the disk, along with slower azimuthal velocities in the plane of the disk.  These differences in velocity distributions indicate that more stars formed in galaxies with jets are formed in regions that have been accelerated outward in the radial and vertical directions and in regions with rotational velocities that have been deccelerated, likely because of the pressurization of the disk from the jet's cocoon.  
Figure \ref{fig:phase_radius_vertical_velocity} shows phase plots of vertical velocity versus disk radius weighted by stellar mass for the disk only simulation at 0.3 Gyr, the disk and jet simulation at 0.3 Gyr, and the disk and jet simulation at 1 Gyr.  We overplot lines on the phase plots showing the $90^{th}$, $75^{th}$, $50^{th}$ (median), $25^{th}$, and $10^{th}$ percentiles of velocity.  The disk only simulation shows very small vertical velocities and small deviations from the average vertical velocity of zero, and also changes very little over the next Gyr of evolution.  The jet simulation data, however, shows much higher vertical velocities and larger vertical velocity dispersions, particularly at 0.03 Gyr, just 5 Myrs of evolution after the end of the jet simulation.  However, the distribution does tighten as the galaxy evolves, particularly at radii $>9$ kpc for $t = 1$ Gyr (lower panel).

\begin{figure}
\includegraphics[width=1.\linewidth,angle=0]{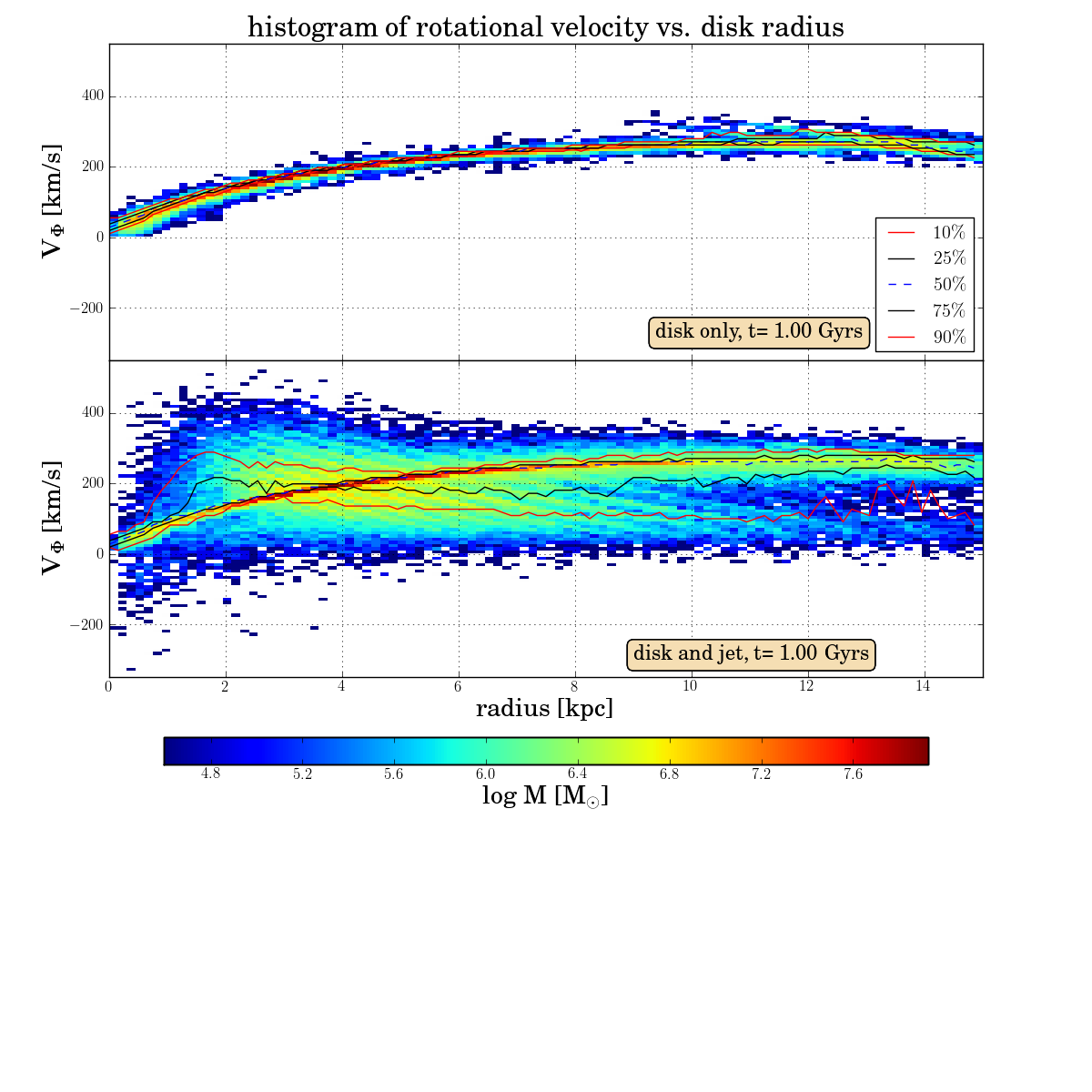}
\vspace{-2.5cm}
\caption{2D histogram of rotational velocity and axial distance for both the disk only simulation and the disk and jet simulation. Both at taken at $t=1$ Gyr and are weighted by stellar mass.}
\label{fig:phase_radius_rotational_velocity}
\end{figure}

Our results show a jet-induced difference in the azimuthal velocity distributions.  We find that in the galaxy with a jet, the rotational velocity of stars within a radius of 4 kpc is greater than in the galaxy without a jet, while the rotational velocity of stars with radii greater than 4 kpc is slower than in the galaxy without the jet.  Figure \ref{fig:phase_radius_rotational_velocity} shows phase plots of rotational velocity versus disk radius.  The simulation without the jet shows only small deviations from the expected rotation curve.  The jet simulation data, however, shows an initial spreading of the expected rotation curve yielding larger distributions of azimuthal velocity.  However, as the stellar motion within the galaxy evolves, stars within a radius of 4 kpc have a higher rotational velocity than the expected rotation curve, while stars further out that 4 kpc have a significantly slower rotational velocity than the disk only simulation.  This signature also lasts throughout the entire 1Gyr evolution.  It is possible that the increase in rotational velocity in smaller radii and the decrease in rotational velocity at larger radii is the result of a larger fraction of elliptical orbits, whose tangential velocities will be greatest at the closest point to the center of the galaxy and smallest at the furthest point form the center of the galaxy.  

\begin{figure}
\includegraphics[width=1.\linewidth,angle=0]{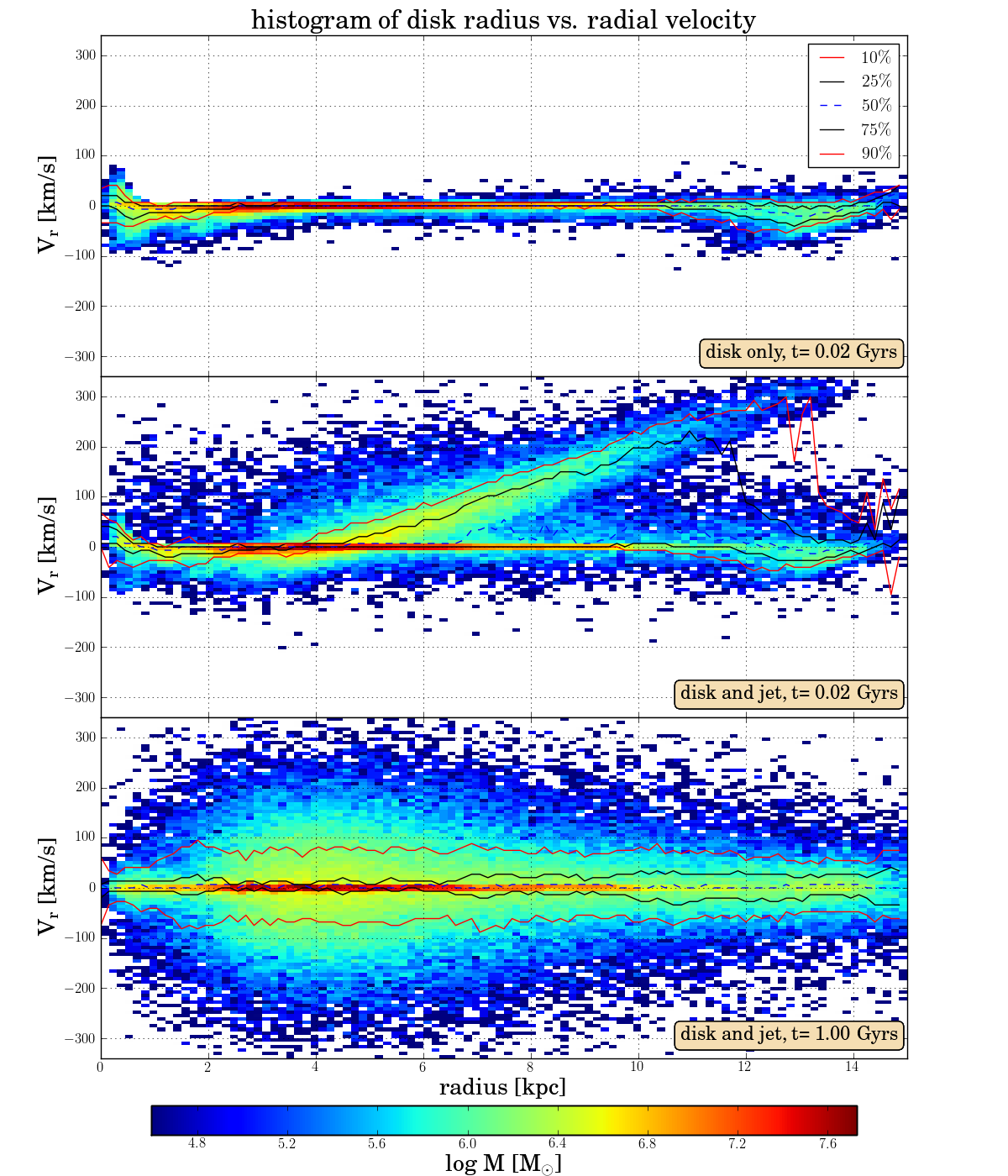}
\caption{2D histogram of radial velocity and axial distance for the disk only simulation at 20 Myr, the disk and jet simulation at 20 Myr, and the disk and jet simulation at 1 Gyr, weighted by stellar mass.}
\label{fig:phase_radius_radial_velocity}
\end{figure}

We also find substantially higher radial velocities in the simulations with the jet.  Figure \ref{fig:phase_radius_radial_velocity} shows phase plots of radial vtelocity versus disk radius for the disk only simulation at 20 Myr, the disk and jet simulation at 20 Myr, and the disk and jet simulation at 1 Gyr.  Without the jet,  only  little to no radial velocity is found for the entire 1 Gyr evolution.  The jet simulation data, however, contains stars that are formed with a large radial boost, particularly stars formed between radii of $\sim 1$ kpc to $\sim 7$ kpc.  This boost is caused by the bow shock emanating from the jet and spreading through the galaxy.  As the galaxy evolves, however, stars with large positive radial velocities are eventually pulled back toward the center of the galaxy.  The positive bias in radial velocity persists through 20 Myr, before these stars are pulled back toward the center of the galaxy.  By around 30 Myr, the distribution of radial velocities evolves toward a large negative bias with stars heading back toward the galaxies center.  
By roughly 0.2 Gyr, however, the fluctuation has stopped, and instead a distribution of radial velocities far larger than in the corresponding snap shot form the disk only simulation is present.  That larger distribution of radial velocities is still present after 1 Gyr of evolution.  The analysis of the radial velocities has important ramifications on assessing the possible origin of HVS because of the large distribution of high radial velocities: roughly $10\%$ of stars formed around a radius of 2 kpc have radial velocities $>300$ km s$^{-1}$. 

%%%%%%%%%%%%%%%%%%%%%%%%%%%%%%%%%%%%%%%%%%%%%%%%%%%%%%%%%%
\section{Escaped Stars and Hyper-Velocity Stars}
\begin{figure}
\begin{center}
\includegraphics[width=1.\linewidth,angle=0]{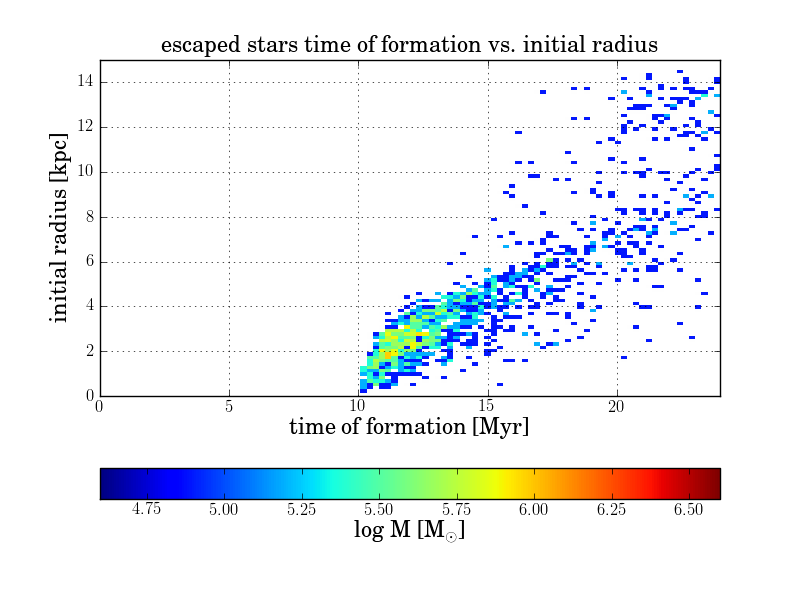}\\
\caption{2D histogram for all escaped stars, showing their time of formation vs. the axial distance of formation, weighted by stellar mass.}
\end{center}
\label{fig:escape_time_formation_radius}
\end{figure}

We define escaped stars as those whose positions exceed spherical radii of 30 kpc.  No stars from the disk only simulations escape the galaxy, and no stars from the jet simulation escape the galaxy until the jet begins at 10 Myr.  Figure~\ref{fig:escape_time_formation_radius} is a phase plot of time of formation vs. initial disk radius of the escaped stars, and clearly shows the bow shock driven by the jet propagates out through the disk and induces the formation of stars in gas clouds with enough momentum to escape the galaxy.  Some of the stars have velocities $>300$ km s$^{-1}$.  We find a decrease in HVS production and a drop in the velocity of those stars after 15 Myr, after the bow shock has passed through the most central part of the disk, consistent with the idea that the shock is generating the HVS.

Some of the stars escape the galaxy with high vertical velocities, moving almost straight off the disk.  These stars were formed closed to the onset of jet activiy.  Other stars escaped the galaxy with high velocities in the plane of the disk, shooting out the edge of the disk.    Most of the escaped stars are formed within a radius of 4 kpc, but some are formed at central distances of $\approx 14$ kpc.  Of the escaped stars formed 10--15 Myr in the simulation, $~75$\% reach a distance from the center of 30 kpc while still remaining within 15 kpc from the jet axis, indicating that most of their velocity is in the vertical direction.  In fact, $~25$\% of these stars formed 10--15 Myr in the simulation have initial vertical velocities greater than 200 km s$^{-1}$.  Of stars formed within 4 kpc distance from the center, about 50\% reach a distance of 30 kpc while still within 10 kpc from jet the axis, and roughly 75\% reach a distance of 30 kpc while still within a 15 kpc distance from the jet axis, indicating again that most of these stars' motion is vertical to the disk.  
An analysis of the time of formation versus the star's radius at formation, as shown in Figure~\ref{fig:escape_time_formation_radius}  traces very closely the pattern of jet induced formation as demonstrated in Figure \ref{fig:phase_time_formation_radius}.  This shows that the jet's bow shock not only induces star formation but also generates high-velocity pockets of star formation.  

%%%%%%%%%%%%%%%%%%%%%%%%%%%%%%%%%%%%%%%%%%%%%%%%%%%%%%%%%%
\section{Parameter Evolution}

To quantify the differing evolution in morphology and velocity distributions, we calculate the sphericity parameter $s$, the triaxiality parameter $T$, the spin parameter $\lambda$, and the velocity anisotropy parameter $\beta$ as a function of time and radius.  We calculate these parameters for the twenty snap shots mentioned in Sect.~\ref{sec:method}. In agreement with the increased distribution in radial velocity distribution, the simulation with the disk and jet has a substantially lower triaxiality for 200--300 Myr after the jet, meaning that the simulations with the jet have more of a flat, oblate disk shape during that time.   
The galaxies with the jet also show lower sphericity for the duration of the 1 Gyr evolution after the jet.  This difference in both parameters indicates that the AGN jet's bow shock creates a more extended stellar structure along the disk of the galaxy.  This conclusion agrees with observationally motivated ideas about AGN feedback `puffing up' the size of massive galaxies at $z \sim 2$ discussed by \citet{Ishibashi}.
\citet{Patel} show that galaxies of a given mass at $z \sim 2$ are more dense than galaxies of similar mass at a redshift of $z \sim 0$, indicating that some mechanism, presumable mergers or AGN feedback, is expanding the size of these galaxies over a period of roughly 10 Gyr.  AGN jets in our simulation trigger star formation in regions with significant radial velocities acting to expand the effective size of the galaxy, particularly at the outer regions (Figs. \ref{fig:time_formation_stellar_mass} and \ref{fig:phase_radius_rotational_velocity}).  Similar observations show that the central regions of the galaxies at $z \sim 2$ are similar to galaxies of the same mass today, indicating the growth in size is occurring at the outer regions \citep{Bezanson,van Dokkum}, a phenomenon we see also in our simulation.

The spin parameter quantifies the coherence of the rotation of the galaxy, and is much higher in the disk only simulations than in the disk and jet simulations.  This is another indicator that the jet disrupts the basic rotation of the galaxy, in agreement with previous analysis of the higher radial and vertical velocity distributions in the simulation with the jet as well as the mostly slower rotational velocity distributions in the simulations with the jet.  The difference in the spin parameter is larger when the calculation includes stars at larger radii, indicating that as an analysis includes more of the outer regions of a galaxy hosting a jet, the less coherent the rotation becomes.  This difference further emphasizes the impact of the jet on the outer regions of the galaxy.  

\begin{figure}
\includegraphics[width=1.\linewidth,angle=0]{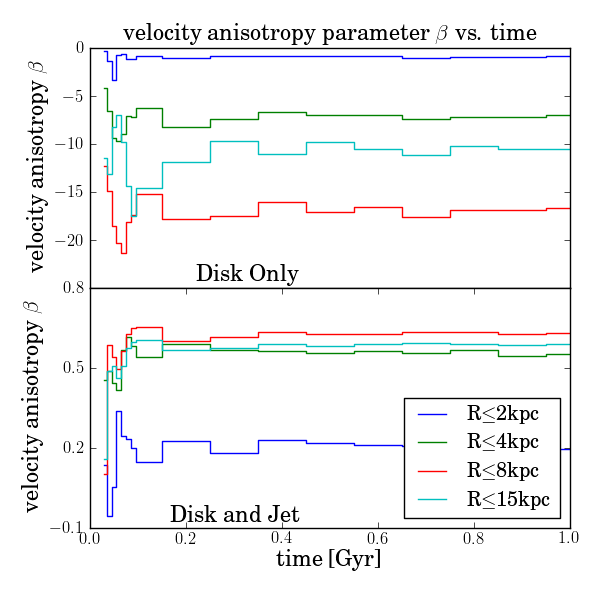} \\
\caption{Velocity anisotropy parameter $\beta$ as function of time for both the disk only simulation and the disk and jet simulation, splitting the stars into four radial regions (axial distance).}
\label{fig:beta}
\end{figure}
Figure \ref{fig:beta} shows the velocity anisotropy parameter $\beta$ over time for both galaxies with and without the jet for four different maximum radii.  The jet causes an extreme difference in the velocity anisotropy parameter $\beta$, and this agrees with our previous analysis of the differing distributions of vertical, radial, and rotational velocity distributions.  The velocity anisotropy parameter for the galaxy with the jet is far closer to 1, between 0.2 and 0.8 for most of the maximum radii, reflecting a much higher bias toward radial motion in the host galaxy than in the disk only simulation, in which $\beta$ is between -20 and -1 for the four maximum radii.  There is more fluctuation in the velocity anisotropy early on for the galaxy with the jet, but the difference in $\beta$ persists for the duration of the Gyr evolution, making increased velocity anisotropy and bias toward radial motion in the host galaxy a lasting signature of past jet activity.  

%%%%%%%%%%%%%%%%%%%%%%%%%%%%%%%%%%%%%%%%%%%%%%%%%%%%%%%%%%
\section{Discussion}

Our analysis identifies several signatures of past AGN jet activity.  The first is the ring of star formation created shortly after the jet becomes active that moves outward radially along the disk starting from a radius of 2 kpc and moving out to a radius of 6 kpc, which corresponds to $\sim 1$ to 4 scale heights of the gaseous disk.  This ring will be one of the most observable signatures to come out of these simulations and in fact may have already been observed in the nearby ($z=0.056$) radio galaxy Cygnus A.  \citet{Jackson} finds a ring of young, blue stars around the center of the galaxy, perpendicular to the direction of the jets.  They contend that the blue light is from young stars and not from scattered quasar radiation. The energy spectrum more closely indicates a possibly instantaneous starburst less than 10 Myr ago.  That time frame would be consistent with the age of AGN activity in the galaxy, also on the order of 10 Myr. Thus, Cygnus A could provide observational evidence for this effect.  

However, because jet activity is so brief on a cosmological time scale, the signatures of jet activity that are still observable long after jet activity ends will be the most useful.  This process also creates a lasting gradient in the stellar populations' time of formation, with the oldest stars having formed in the center of the galaxy and the average age of stellar populations becoming younger as one moves out along the disk.  This also generates the formation of young stars along the outer edges of the galaxy, far from the center.  In fact, not only do \citet{Juneau13} and \citet{Fang12} find possible AGN-triggered star formation, \citet{Fang12} finds rings of star formation in galaxies with AGN but with no current jet activity, consistent with an inside-out positive feedback model as we see in our simulations.  The gradient in the age of stellar populations from the center of the galaxy outward could be a strong signature of jet-triggered star formation.  Furthermore, \citet{Juneau13} finds extended star formation at high radii in galaxies that host AGN.  While these galaxies do not host current jet activity, star formation at the outer parts of the galaxy is consistent with our results.

Our analysis, also shows much wider distributions for radial, vertical and azimuthal velocities.  These increased distributions all persist up to a Gyr after the jet stops.  Of course, our simulations only show these signatures for the stars formed during these 25 Myr and not for stars formed for the next Gyr after, which would effect how observable these effects would be.  These increased distributions also create galaxies with higher velocity anisotropy parameters and lower spin parameters, indicating more random and less coherent orbits within the galaxy, with much higher biases toward radial motions in galaxies that hosted jets in the past. These characteristics all may endure for a Gyr after the jet ends, depending on the stellar mass originating from the jet-induced star formation.

In agreement with the increased distribution of radial velocities, in particular the positive bias in radial velocity with the initialization of the jet, we find that high-velocity stars and escaped stars are formed in the exact same pattern as the previously discussed rings of star formation, as characterized by their radius of formation and time of formation.  Therefore, a distribution of escaped stars that matches that pattern could also indicate past jet activity, and perhaps past jet induced star formation.  

While the Milky Way is a $z=0$, barred spiral galaxy, and the galaxies we simulate are high-redshift and gas-rich, it is possible that some of the signatures of AGN jet activity in the galaxies we simulate may also be present in the Milky Way.  These results may also be examined with the data of  the upcoming space astrometry mission Gaia, which will accurately measure the positions, proper motions, and velocities of a billion stars in the Milky Way.  These comparisons would have to be between general results, such as a noticeably positive bias in the radial velocity distribution. \citet{Deason} has already measured the velocity anisotropy parameter of some halo stars in the Milky Way.  As similar analyses are published, comparison with a theoretical approach as described here will yield productive results for the role of AGN jets possibly in the Milky Way and more generally  in galaxy formation and evolution.

%%%%%%%%%%%%%%%%%%%%%%%%%%%%%%%%%%%%%%%%%%%%%%%%%%%%%%%%%%
\acknowledgments

ZD was supported by a Centre for Cosmological Studies Balzan Fellowship.  
VG was supported by the Sonderforschungsbereich SFB 881 ``The Milky Way System'' (subproject B4) of the German Research Foundation (DFG).
The research of JS has been supported at IAP by  the ERC project  267117 (DARK) hosted by Universit\'e Pierre et Marie Curie - Paris 6   and at JHU by NSF grant OIA-1124403.

%%%%%%%%%%%%%%%%%%%%%%%%%%%%%%%%%%%%%%%%%%%%%%%%%%%%%%%%%%

\end{document}